\documentclass[%
 aip,
 amsmath,amssymb,
 reprint,%
]{revtex4-1}

\usepackage{graphicx}
\usepackage{dcolumn}
\usepackage{bm}

\usepackage[utf8]{inputenc}
\usepackage[T1]{fontenc}
\usepackage{mathptmx}
\usepackage{etoolbox}
\usepackage{amsmath}
\usepackage{braket}
\usepackage{color}
\usepackage{ulem}

\definecolor{darkgreen}{RGB}{40,110,5}

\usepackage{ulem}  
\definecolor{orange}{rgb}{1,0.5,0}

\newcommand{\bfepsilon}{\mathbf{\epsilon}\hspace{-4.5pt}\mathbf{\epsilon}}
\newcommand{\bfmu}{\mathbf{\mu}\hspace{-7.4pt}\mathbf{\mu}}

\makeatletter
\def\@email#1#2{%
 \endgroup
 \patchcmd{\titleblock@produce}
  {\frontmatter@RRAPformat}
  {\frontmatter@RRAPformat{\produce@RRAP{*#1\href{mailto:#2}{#2}}}\frontmatter@RRAPformat}
  {}{}
}%
\makeatother
\begin{document}

\preprint{AIP/123-QED}

\title{Simulating optical linear absorption for mesoscale molecular aggregates: \\an adaptive hierarchy of pure states approach}
\author{Tarun Gera}
\affiliation{Department of Chemistry, Southern Methodist University, PO Box 750314, Dallas, TX, USA}
\author{Lipeng Chen}
\affiliation{Max Planck Institute for the Physics of Complex Systems, N\"{o}thnitzer Str 38, Dresden, Germany}
\author{Alex Eisfeld}
\affiliation{Max Planck Institute for the Physics of Complex Systems, N\"{o}thnitzer Str 38, Dresden, Germany}
\author{Jeffrey R. Reimers}
\affiliation{International Centre for Quantum and Molecular Structures and the School of Physics, Shanghai University, 200444, Shanghai, China}
\affiliation{School of Mathematical and Physical Sciences, University of technology Sydney, NSW 2007, Australia}
\author{Elliot J. Taffet}
\affiliation{Department of Chemistry, Southern Methodist University, PO Box 750314, Dallas, TX, USA}
\author{Doran I. G. B. Raccah}%
\affiliation{Department of Chemistry, Southern Methodist University, PO Box 750314, Dallas, TX, USA}
\email{doranb@smu.edu}

\date{\today}

\begin{abstract}
In this paper, we present a new method for calculating linear absorption spectra for large molecular aggregates, called dyadic adaptive HOPS (DadHOPS). 
This method combines the adaptive HOPS (adHOPS) framework, which uses locality to improve computational scaling, with the dyadic HOPS method previously developed to calculate linear and non-linear spectroscopic signals. 
To construct a local representation of dyadic HOPS, we introduce an initial state decomposition which reconstructs the linear absorption spectra from a sum over locally excited initial conditions. 
We demonstrate the sum over initial conditions can be efficiently Monte Carlo sampled, that the corresponding calculations achieve size-invariant (i.e. $\mathcal{O}(1)$) scaling for sufficiently large aggregates, and that it allows for the trivial inclusion of static disorder in the Hamiltonian. 
We present calculations on the photosystem I core complex to explore the behavior of the initial state decomposition in complex molecular aggregates, and proof-of-concept DadHOPS calculations on an artificial molecular aggregate inspired by perylene bis-imide. 
\end{abstract}

\maketitle

\section{Introduction}
Optical spectroscopy provides essential insight into excited-state processes of molecular aggregates ranging from artificial assemblies in solution to photosynthetic pigment-protein complexes. \cite{van2000photosynthetic,may2000charge, Abramavicius2009,Chen2015, mukamel1995principles}
The interpretation of spectroscopic observables in the presence of both structural heterogeneity and broad homogeneous lineshapes often requires detailed numerical simulations.
However, established quantum dynamics methods can struggle with the combination of strong electron-vibrational coupling and  spatially-extended structures characteristic of molecular aggregates.\cite{Gelin2022}
As a result, the development of efficient algorithms for simulating time-resolved optical spectroscopy measurements of extended molecular aggregates remains a persistent challenge.

From a molecular perspective, simulating optical spectroscopy requires solving the time-evolution of nuclear wave packets on electronic potential energy surfaces. A variety of methods have been developed to solve this time evolution within a formally exact framework such as  multiconfiguration time-dependent Hartree (MCTDH), \cite{Beck2000,Burghardt2003,Richings2015} multilayer MCTDH (ML-MCTDH),\cite{Wang2003,Wang2015,Manthe2008,Vendrell2011} multi-configuration Ehrenfest,\cite{Shalashilin2009,Shalashilin2010,Chen2019} and \textit{ab initio} multiple spawning.\cite{Ben1998,Ben2000} 
However, the simultaneous description of electronic, intra-molecular vibrational, and environmental degrees of freedom on equal footing introduces intractable computational scaling for large molecular aggregates; a problem colloquially known as the curse of dimensionality.  

Open-quantum system approaches provide a powerful coarse-graining where the relevant electronic degrees of freedom are time-evolved in a reduced density matrix while coupled to an effective thermal environment. 
Most commonly, an open-quantum system Hamiltonian incorporates a linear coupling to a infinite collection of harmonic oscillators parameterized to mimic the influence of both the intramolecular vibrations and the surrounding (condensed phase) environment within a linear response approximation.  
Within this framework, several formally exact methods are available for modeling molecular excitons including Hierarchical equations of motion (HEOM),  \cite{Tanimura2006,tanimura2020numerically} quasi-adibatic path integrals,\cite{Makri1992,makri1995} time-dependent density matrix renormalization group theory (TD-DMRG),\cite{Marston2002,White2004} time-evolving density operator with orthogonal polynomials,\cite{prior2010efficient,Tamascelli2019} and the multi-D1 Davydov ansatz.\cite{Chen2017,Chen2018} 
While these methods are capable of scaling to much larger aggregates than are usually achievable with wave function propagation techniques, they still suffer from effectively exponential scaling with the number of electronic states (i.e. molecules) included in the calculations. 
Recent developments in reduced scaling techniques - including modular path integrals\cite{makriCommunicationModularPath2018,makriModularPathIntegral2018} and tensor contraction\cite{borrelliExpandingRangeHierarchical2021,yanEfficientPropagationHierarchical2021,somozaDissipationAssistedMatrixProduct2019,tamascelliEfficientSimulationFiniteTemperature2019,strathearnEfficientNonMarkovianQuantum2018b} - suggest new paths forward may continue to extend the size of molecular aggregates that are tractable with these methods. 

Recently, there has been a growing interest in solving open-quantum systems using stochastic wave functions. In these approaches, the reduced density matrix is unravelled into an ensemble of wave functions that can be time-evolved independently. In this context, the non-Markovian quantum state diffusion (NMQSD) equation \cite{Diosi1997,Strunz1998PRA} provides a formally exact solution to the dynamics. The hierarchy of pure states (HOPS) is a numerically convenient formulation of the NMQSD equations \cite{Suess2014,Hartmann2017,Gerhard2015JCP} which has been recently extended into the dyadic HOPS equation to simulate both linear \cite{Chen2022} and non-linear\cite{Eisfeld2022} optical spectroscopy. Dyadic HOPS propagates the bra- and ket-side of the reduced density matrix separately according to the HOPS equation and provides a clear connection to the non-linear response function formalism.\cite{mukamel1995principles}

Dyadic HOPS is limited to small molecular aggregates by the same poor computational scaling that plagues HOPS. 
The recent development of adaptive HOPS (adHOPS), which leverages the locality of physical wave functions to achieve computational costs that stop increasing with system-size for sufficiently large aggregates (i.e.~$\mathcal{O}(1)$ scaling),\cite{Varvelo2021} raises the possibility of dramatically expanding the reach of dyadic HOPS within an adaptive framework. In this paper, we demonstrate dyadic adaptive HOPS (DadHOPS) calculations of linear absorption for large molecular aggregates. 
We present an initial state decomposition for the dipole auto-correlation function that characterizes linear absorption which provides a local construction of the spectroscopic observable. 
By combining DadHOPS with an initial state decomposition, we are able to perform size-invariant scaling simulations of linear absorption spectra for mesoscale molecular aggregates and easily incorporate the influence of static disorder in the system Hamiltonian.

This paper is organized as follows: In Sec.~\ref{Sec:TB}, we provide a theoretical background for the readers. We discuss the Hamiltonian for an open quantum system followed by a brief introduction for the HOPS and dyadic HOPS methods. In Sec.~\ref{Sec:DyadicAdHOPS}, we describe an algorithm to connect adaptivity with dyadic HOPS method and provide conditions for relative error bounds that are necessary for DadHOPS. In Sec.~\ref{Sec:MC}, we apply the initial state decomposition method to decompose the dipole autocorrelation function into a sum over local correlation functions.
We discuss an effective method for Monte Carlo sampling over stochastic noise trajectories and initial states, with both dyadic HOPS and DadHOPS. Using 4-site and 12-site chains as examples we provide proof-of-concept calculations for the applicability of the DadHOPS method. In Sec.~\ref{Sec:Application}, we apply the initial state decomposition to photosystem I and then we apply DadHOPS to a realistic model for artificial molecular aggregates inspired by Perylene bis-imide (PBI). Finally, we conclude in Sec. \ref{Sec:Conclusions} with a summary and a brief outlook. In Appendix A, we provide a detailed derivation of the dipole autocorrelation function decomposition into a sum of local correlation functions based on a generalized initial state decomposition.

\section{Theoretical Background} \label{Sec:TB}

\subsection{Hamiltonian} \label{Sec:Hamiltonian}
We model molecular aggregates consisting of $N$ chromophores with an open-quantum system Hamiltonian 
\begin{equation}\label{eq:Htot}
\hat{H}=\hat{H}_{\mathrm{S}}+\hat{H}_{\mathrm{B}}+\hat{H}_{\mathrm{SB}}
\end{equation}
where the system Hamiltonian  
\begin{equation}
\label{eq:H_sys}
\hat{H}_{\mathrm{S}}=E_\mathrm{g}|\mathrm{g}\rangle\langle{\mathrm{g}}|+\sum_{n=1}^{\mathrm{N}}E_n|n\rangle\langle{n}|+{\sum_{n=1}^N}\sum_{m\neq n}^{N}V_{nm}|n\rangle\langle{m}|
\end{equation}
is composed of a shared electronic ground state, electronic excited states with vertical excitation energy ($E_n$), and an electronic coupling between pigments $V_{nm}$. The electronic states of each pigment are linearly coupled 
\begin{equation}
\hat{H}_{\mathrm{SB}} = -\sum_{n=1}^N \hat{L}_n \sum_qg_{nq}\left(\hat{b}_{nq}^{\dagger}+\hat{b}_{nq}\right) 
\end{equation}
to an independent harmonic reservoir 
\begin{equation}
    \hat{H}_{\mathrm{B}}=\sum_{n=1}^N\sum_q  \hbar\omega_{nq}\hat{b}_{nq}^{\dagger}\hat{b}_{nq}
\end{equation}
with a system-bath coupling operator $\hat{L}_n= \ket{n}\bra{n}$. The influence of the vibrational modes on the dynamics of the electronic system is described by the bath-correlation function
\begin{equation}
\begin{aligned}
\alpha_n(\tau)=\int_0^{\infty}\!\!\!\mathrm{d}\omega\, J_n(\omega)\Big( \coth\big(\frac{\beta \hbar\omega}{2}\big) \cos (\omega \tau) -i \sin(\omega \tau)\Big)
\end{aligned}
\end{equation}
that contains the spectral density $J_n(\omega) = \sum_{q}|g_{nq}|^2\delta(\omega-\omega_{nq})$ and the inverse temperature $\beta=\frac{1}{\textrm{k}_B T}$. We decompose the bath correlation function into a sum of exponentials indexed by $j_n$
\begin{equation}
\label{eq:alpha_dl_coarse}
    \alpha_n(t) = \sum_{j_n} g_{j_n} e^{-\gamma_{j_n} t/\hbar} 
\end{equation}
where $g_{j_n}$ and $\gamma_{j_n}$ are, in general, complex valued. 

\subsection{Light-Matter Interaction}
The light matter interaction is described in terms of the scalar, collective dipole moment operator 
\begin{equation}
\label{eq:collective_Transition_Operator}
\hat{\mu}_\mathrm{eff}=\sum_{n=1}^N (\bfmu_n\cdot \bfepsilon) \,|n\rangle\langle{\mathrm{g}}|+ h.c.,
\end{equation}
which is defined by the polarization of the incident electric field ($\bfepsilon$) and the transition dipole operator the individual pigments ($\bfmu_n$). The action of the collective dipole moment operator on the ground state of the aggregate is to excite the superposition state
\begin{equation}
\label{eq:psi_ex}
    \ket{\psi_\mathrm{ex}} = \frac{1}{{\mu}_\mathrm{tot}} \hat{\mu}_\mathrm{eff} \ket{\mathrm{g}}
\end{equation}
where $\mu_\mathrm{tot}= \sqrt{\sum_{n=1}^N (\bfmu_n \cdot \bfepsilon)^2 }$. 

The linear absorption spectrum is given by the half-sided Fourier transform of the dipole auto-correlation function 
\begin{eqnarray}\label{eq:Ct_def}
C(t)
=\mathrm{Tr}\left\lbrace\hat{\mu}_{\mathrm{eff}}\,{e}^{-i\hat{H}t/\hbar}\big(\hat{\mu}_\mathrm{eff}{|}\mathrm{g}\rangle\langle{\mathrm{g}}|\otimes\hat{\rho}_{\mathrm{B}}\big) e^{i\hat{H}t/\hbar}\right\rbrace
\end{eqnarray}
where $\hat{\rho}_0=|\mathrm{g}\rangle\langle{\mathrm{g}}|\otimes\hat{\rho}_{\mathrm{B}}$ is a factorized initial total density matrix with $\hat{\rho}_{\mathrm{B}}=e^{-\beta{\hat{H}}_{\mathrm{B}}}/\mathrm{Tr}_{\mathrm{B}}\left\lbrace{e}^{-\beta{\hat{H}}_{\mathrm{B}}}\right\rbrace$ being the density matrix of the thermal bath. 


\subsection{Hierarchy of Pure States (HOPS)}
The Hierarchy of Pure States (HOPS) is a formally exact solution to the open-quantum system Hamiltonian presented in Section \ref{Sec:Hamiltonian}. Using HOPS, time evolution of the system reduced density matrix, starting from a separable initial state ($\hat{\Phi}(0) = \ket{\psi(0)}\bra{\psi(0)} \bigotimes \hat{\rho}_B$), can be described in terms of an ensemble average ($\mathbb{E}[\cdot]$) over stochastic wave functions 
\begin{equation}
    \rho(t) = \mathbb{E}\Bigg[\frac{\ket{\psi^{(\vec{0})}(t; \mathbf{z}^*)} \bra{\psi^{(\vec{0})}(t; \mathbf{z}^*)}}{\braket{\psi^{(\vec{0})}(t; \mathbf{z}^*) \vert \psi^{(\vec{0})}(t; \mathbf{z}^*)}}\Bigg] 
\end{equation}
where $\mathbf{z}^*$ is a stochastic trajectory with components associated with individual thermal environments $z_{n,t}$ defined by $\mathbb{E}[z_{n,t}]=0$,  $\mathbb{E}[z_{n,t} z_{n,s}]=0$, and $\mathbb{E}[z^*_{n,t} z_{n,s}]=\alpha_n(t-s)$. The time-evolution of the system wave functions can be calculated using the non-linear HOPS equation
\begin{flalign}
\begin{aligned}
\label{eq:NormNonLinearHops}
\hbar \partial_t \vert \psi^{(\Vec{k})}(t; \mathbf{z}^*) \rangle 
=  &\big(-i\hat{H}_S - \Vec{k} \cdot \Vec{\gamma} + \sum_{n} \hat{L}_{n} (z^*_{n,t}+ \xi_{n,t})\big)\vert \psi^{(\Vec{k})}(t; \mathbf{z}^*) \rangle \\ 
+ &\sum_{n,j_n} k_{j_n} \gamma_{j_n} \hat{L}_{n}  \vert \psi^{(\Vec{k} -\Vec{e}_{j_n})}(t; \mathbf{z}^*) \rangle \\
- &\sum_{n,j_n} (\frac{g_{j_n}}{\gamma_{j_n}})(\hat{L}^{\dagger}_{n} - \langle\hat{L}^{\dagger}_{n}\rangle_{t}) \vert \psi^{(\Vec{k}+\Vec{e}_{j_n})}(t; \mathbf{z}^*)\rangle, 
\end{aligned}
\end{flalign}
where the physical wave function is given by $\vert \psi^{(\vec{0})}(t; \mathbf{z}^*) \rangle$, the auxiliary wave functions are indexed by a vector $\Vec{k}$, $\Vec{\gamma}$ is the vector of correlation function exponents ($\gamma_n$), and 
\begin{equation}
    \xi_{t,n} = \frac{1}{\hbar}\int_{0}^{t} ds \alpha^{*}_{n}(t-s) \braket{L^{\dagger}_{n}}_s
\end{equation}
is a memory term that causes a drift in the effective noise. The expectation value of the system-bath coupling operator is 
\begin{equation}
    \langle\hat{L}^{\dagger}_{n}\rangle_{t} = \frac{\langle \psi^{(\vec{0})}(t; \mathbf{z}^*) \vert \hat{L}^{\dagger}_{n}\vert \psi^{(\vec{0})}(t; \mathbf{z}^*) \rangle}{\braket{\psi^{(\vec{0})}(t; \mathbf{z}^*) \vert \psi^{(\vec{0})}(t; \mathbf{z}^*)}}.
\end{equation}
We limit the hierarchy to a finite depth $k_{max}$ using the triangular truncation condition where we only include auxiliary wave functions satisfying the condition $\{\Vec{k} \in \mathbb{A}:  \sum_i k_i \leq k_{max}\}$, though other static filtering approaches have been explored previously. \cite{Zhang2018}

\subsection{Dyadic HOPS}
Following the introduction of the pure state decomposition by Hartmann et al.\cite{Hartmann2021} and the subsequent work of Chen et al.,\cite{Chen2022} the dipole autocorrelation function that defines the linear absorption spectrum can be written as
\begin{equation}
\label{eq:C(t)_diag}
C(t)= \mu_\mathrm{tot} \sum_{\eta\in \{\pm 1, \pm i\}} \frac{\eta}{2} \,\mathbb{E} [\bra{v_\eta(t;\mathbf{z}^*)} \hat\mu_\mathrm{eff} \ket{v_\eta(t;\mathbf{z}^*)}  ]
\end{equation}
where $\ket{v_\eta(t;\mathbf{z}^*)}$ is the initial state 
\begin{equation}
    \ket{v_\eta}=\frac{1}{\sqrt{2}} (\ket{\psi_\mathrm{ex}}+ \eta\ket{\mathrm{g}})
\end{equation}
time-evolved in the full Hilbert space and $\eta$ is the index of the four pure state initial conditions. This correlation function reduces (accounting for cancellation between $\eta$ terms), to 
\begin{equation}
\label{eq:C(t)_psd}
C(t)=  \mu_\mathrm{tot}^2 \mathbb{E}[\braket{\psi_\mathrm{ex}|\psi_\mathrm{ex}(t;\mathbf{z}^*)}]e^{i E_{\mathrm{g}} t/\hbar}.
\end{equation}
As written, Eq.~\eqref{eq:C(t)_psd} requires the system wave functions to be propagated in the Hilbert space containing both the electronic ground and excited state. The dyadic HOPS equation, introduced in Ref.~\onlinecite{Chen2022}, implicitly accounts for the influence of the ground state but only time evolves the excited states. The dyadic HOPS equation is equivalent to Eq.~\eqref{eq:NormNonLinearHops} except that the expectation value of a system-bath coupling operator is given by 
\begin{equation}
\label{eq: dyadicHOPS_renormalization}
    \langle\hat{L}^{\dagger}_{n}\rangle_{t} = \frac{\langle \psi^{(\vec{0})}(t; \mathbf{z}^*) \vert \hat{L}^{\dagger}_{n}\vert \psi^{(\vec{0})}(t; \mathbf{z}^*) \rangle}{\braket{\psi^{(\vec{0})}(t; \mathbf{z}^*) \vert \psi^{(\vec{0})}(t; \mathbf{z}^*)}+1},
\end{equation}
and the dipole autocorrelation function becomes 
\begin{equation}
\label{eq:C(t)_final_normalized}
C(t)=  \mu_\mathrm{tot}^2 \mathbb{E}\Bigg[\frac{\braket{\psi_\mathrm{ex} | \psi_\mathrm{ex}(t;\mathbf{z}^*)}}{\frac{1}{2}\left(||\psi_\mathrm{ex}(t;\mathbf{z}^*)||^2 + 1\right)}\Bigg]e^{i E_{\mathrm{g}} t/\hbar}.
\end{equation}

\section{Dyadic adaptive HOPS} \label{Sec:DyadicAdHOPS}
Here, we present an algorithm for constructing an adaptive basis during a dyadic HOPS trajectory that can substantially reduce computational requirements when simulating large systems. The algorithm presented here builds on previous work implementing adaptive HOPS for a density matrix calculation. \cite{Varvelo2021} The basic physical picture is that HOPS trajectories are localized by the presence of their thermal environments (i.e. bath Hamiltonians) and, as a result, the time-evolution can be described by a local basis that moves with the trajectory. To ensure the adaptive calculations retain the formally exact structure of HOPS, the local basis must guarantee a controllable error bound on the calculation. 


\begin{figure}
\includegraphics[width=0.5\textwidth]{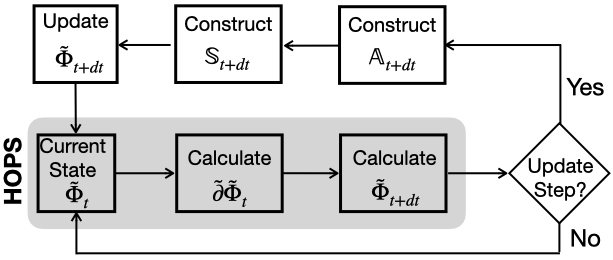}
\caption{\label{fig:AdaptiveAlgorithm} Algorithm for adHOPS }
\end{figure}

Previously, Ref. \onlinecite{Varvelo2021} has demonstrated that it is possible to construct an adaptive HOPS algorithm where computational expense does not scale with system size for sufficiently large aggregates (i.e.\ $\mathcal{O}(1)$ scaling). Fig.~\ref{fig:AdaptiveAlgorithm} presents a sketch of this algorithm, where the local basis is constructed as a direct sum ($\mathbb{A}_t \bigoplus \mathbb{S}_t$) of set of auxiliary wave functions ($\mathbb{A}_t$, called the `auxiliary basis') and the set of molecular states ($\mathbb{S}_t$, called the `state basis'). Every time point, we build a new auxiliary and state basis that ensures the difference between the full derivative of all wave functions (represented by $\partial \Phi$) and the derivative constructed in the reduced basis $\tilde{\partial} \tilde{\Phi}$ is less than a user specified threshold $\delta$ (i.e.\ $\vert \vert{\partial \Phi - \tilde{\partial} \tilde{\Phi}}\vert\vert<\delta$).
For convenience, we split the user specified derivative error bound ($\delta = \sqrt{\delta_A^2 + \delta_S^2}$) into two components, the auxiliary bound ($\delta_{A}$) and the state bound ($\delta_S$), which allow us to independently control the precision we require when constructing the auxiliary and state bases, respectively.  

Extending the algorithm to dyadic adaptive HOPS (DadHOPS) requires a generalization to account for the fact that the norm of the physical wave function ($\ket{\psi^{(\vec{0})}(t; \mathbf{z}^*)}$) is not conserved during the time evolution of the trajectory.
To ensure a consistent accuracy across the trajectory, we use relative error bounds $\left( \Delta_{A/S}(t) = \delta_{A/S} \cdot \vert \vert \psi^{(\vec{0})}(t; \mathbf{z}^*) \vert \vert \right)$ that evolves in time with the norm of the physical wave function. 

All calculations reported here are run with MesoHOPS v1.2.1 \cite{MesoHOPS}  \footnote{The most recent version of MesoHOPS can be found here: https://github.com/MesoscienceLab/mesohops}

\section{Monte Carlo Sampling Local Trajectories} \label{Sec:MC}
Here, we present the combinations of the DadHOPS algorithm with a local representation of the dipole autocorrelation function to enable efficient simulation of absorption spectra for mesoscale molecular aggregates. 
For the DadHOPS algorithm to be efficient the spatial extent of delocalization must be substantially smaller than the full size of the molecular aggregate. Simultaneously, implemented directly, the dyadic HOPS expression for the dipole autocorrelation function has an initial condition that can be arbitrarily delocalized. 
In the following, first, we demonstrate  that the dipole autocorrelation function can be decomposed into local initial conditions. 
Second, we demonstrate that the total correlation function can be reproduced by a simultaneous Monte Carlo sampling of noise trajectories ($\mathbf{z}^*_t$) and local initial conditions. Finally, we combine the initial state decomposition with the DadHOPS algorithm to demonstrate a local construction of a linear absorption spectrum that can achieve size-invariant (i.e. $\mathcal{O}(1)$) scaling.  

As a model system, we consider a linear chain of $N_\mathrm{pig}$ molecules with parallel dipole moments. The electronic coupling is assumed to be nearest neighbor ($V = -100 \textrm{ cm}^{-1}$) and we describe the bath correlation function as 
\begin{equation}
    \alpha_n(t) = \frac{i \lambda \Gamma_-}{\beta \nu} e^{-\Gamma_+ t/\hbar} + \frac{-i \lambda \Gamma_+}{\beta \nu} e^{-\Gamma_- t/\hbar}
\end{equation}
which is a high-temperature approximation to the spectral density described in Ref. \onlinecite{Ishizaki2020}. For these calculations, we use $\lambda = 35 \textrm{ cm}^{-1}$, $\Gamma=50 \textrm{ cm}^{-1}$, $\nu = 10 \textrm{ cm}^{-1}$, $\Gamma_{\pm} = \Gamma \pm i\nu$, and $T=295$ K.

\subsection{Initial state decomposition}
We can expand the collective dipole operator 
\begin{equation}
    \hat{\mu}_{\mathrm{eff}} = \sum_\mathbf{d} A_\mathbf{d} \hat{\sigma}_\mathbf{d} 
\end{equation}
into a sum over local excitation operators acting on a cluster of pigments ($\mathbf{d}$)
\begin{equation}
    \hat{\sigma}_\mathbf{d} =  \sum_{d \in \mathbf{d}} \frac{\boldsymbol{\mu}_{d}\cdot\bfepsilon}{A_\mathbf{d}} \left(\ket{d}\bra{g} + \ket{g}\bra{d}\right)
\end{equation}
where 
\begin{equation}
    \label{eq:mu_d}
    A_\mathbf{d} = \sqrt{\sum_{d\in\mathbf{d}} (\bfmu_{d} \cdot \bfepsilon)^2 }.
\end{equation}
These equations assume that the pigment clusters represent a partition of the set of all pigments (i.e. they are disjoint and their union contains all pigments), though a more general construction is described in Appendix A. 
Inserting this expansion into the definition of the dipole autocorrelation function (Eq.~\eqref{eq:Ct_def}) for the first interaction with the electric field we find 
\begin{equation}
    C(t) = \sum_\mathbf{d} A_\mathbf{d} C_\mathbf{d}(t)
\end{equation}
where 
\begin{equation}
    C_\mathbf{d}(t) = \mathrm{Tr}\left\lbrace\hat{\mu}_{\mathrm{eff}}\,{e}^{-i\hat{H}t/\hbar}\big( \hat{\sigma}_\mathbf{d}{|}\mathrm{g}\rangle\langle{\mathrm{g}}|\otimes\hat{\rho}_{\mathrm{B}}\big) e^{i\hat{H}t/\hbar}\right\rbrace.
\end{equation}
In Appendix A, we prove that the local correlation function contribution can be calculated as
\begin{equation}
\label{eq:C(t)_local_dyadic_cluster}
C_\mathbf{d}(t)= \mu_\mathrm{tot} \, \mathbb{E} \Bigg[ \frac{\braket{\psi_\mathrm{ex}|\psi_\mathbf{d}(t; \mathbf{z}^*)}}{\frac{1}{2}(\vert \vert  \psi_\mathbf{d}(t; \mathbf{z}^*) \vert \vert^2+1)} \Bigg] e^{i E_{\mathrm{g}} t/\hbar}
\end{equation}
where $\psi_\mathbf{d}(t; \mathbf{z}^*)$ is the initial state 
\begin{equation}
    \ket{\psi_\mathbf{d}} = \hat{\sigma}_\mathbf{d} \ket{g}
\end{equation}
time-evolved according to the dyadic HOPS equation. In the special case where each cluster contains a single pigment ($n$), this equation reduces to 
\begin{equation}
\label{eq:C(t)_final_pigment}
C_n(t)= \mu_{\mathrm{tot}}\, \mathbb{E}\Big[\frac{\braket{\psi_\mathrm{ex} \vert n(t;\mathbf{z}^*)}}{\frac{1}{2}\left(||n(t;\mathbf{z}^*)||^2 + 1\right)}\Big]e^{i E_{\mathrm{g}} t/\hbar}
\end{equation}
where $\hat{\sigma}_n = (\ket{n}\bra{g} + \ket{g}\bra{n}), \,\,\, A_n = \bfmu_{n} \cdot \bfepsilon,$
and $\ket{n(t;\mathbf{z}^*)}$ is a single-site initial excitation time-evolved according to the dyadic HOPS equations. 

\begin{figure}
\includegraphics{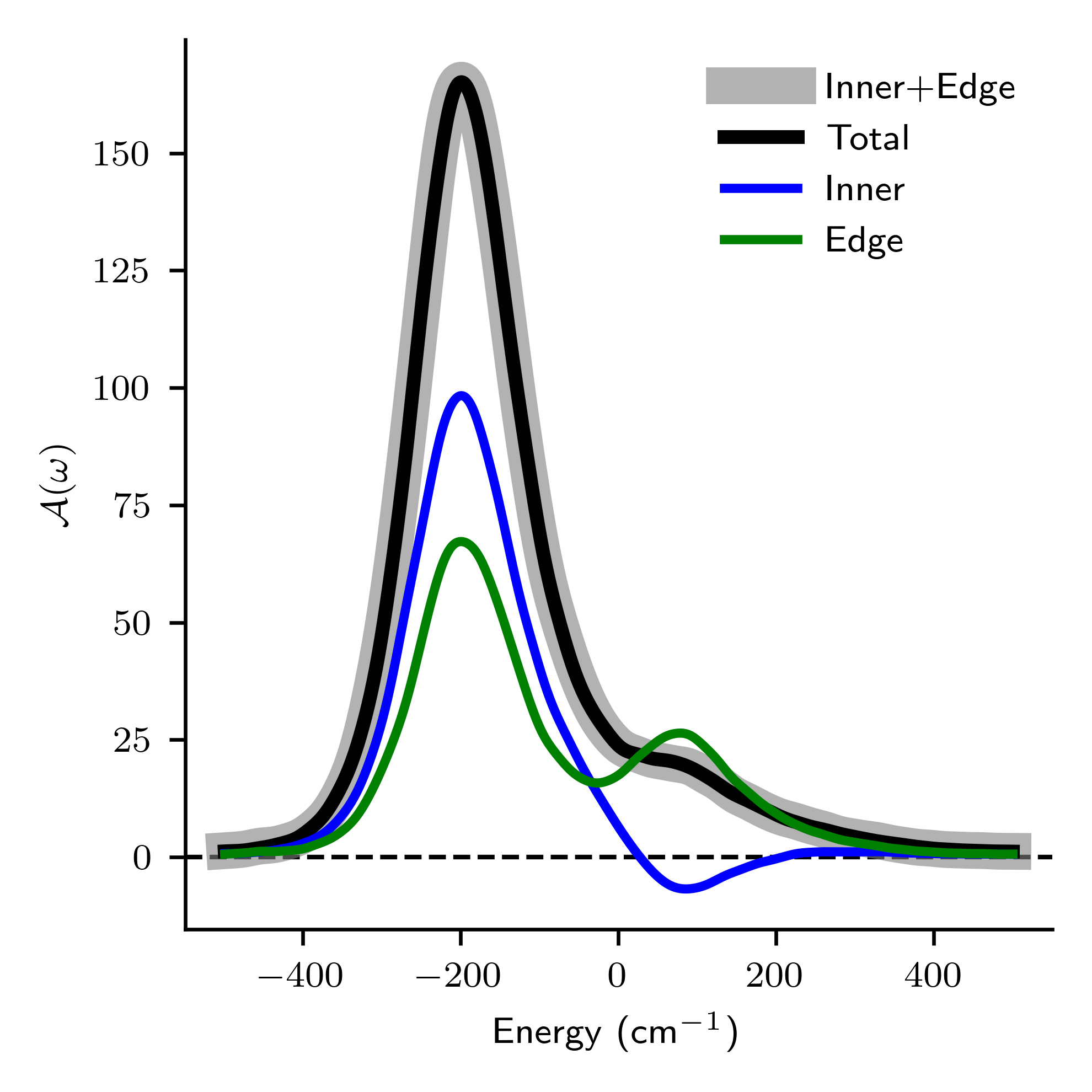}
 \caption{\label{fig:ISD_spectra}  The initial state decomposition of a linear absorption spectrum. The total absorption spectrum (black) calculated for the 4-site chain agrees with the sum (grey) of the edge (green) and inner (blue) spectral contributions. 
 Parameters: $V=-100 \textrm{ cm}^{-1}$, $\lambda = 35 \textrm{ cm}^{-1}$, $\Gamma=50 \textrm{ cm}^{-1}$, $\nu = 10 \textrm{ cm}^{-1}$, $T=295 \textrm{ K}$, and $k_{\mathrm{max}}=5$.}   
 \end{figure}

\paragraph{Example:}
To provide insight into the initial state decomposition and to show that it reproduces the total absorption spectrum, we perform calculations on a four site ($N_\mathrm{pig} = 4$) homogeneous chain ($E_n = 0$). In this case, the total correlation function can be decomposed into a sum of four single-site initial conditions ($C_n(t)$)
\begin{equation}
    C(t) = \mu_0 (C_1(t) + C_2(t) + C_3(t) + C_4(t))
\end{equation}
and $\mu_0 = \bfmu_n \cdot \bfepsilon$. The symmetry of the Hamiltonian gives rise to only two unique initial conditions 
\begin{align}
    C_{\mathrm{edge}}(t) &= \mu_0 (C_1(t) + C_4(t)) = 2 \mu_0 C_1(t) \\
    C_{\mathrm{inner}}(t) &= \mu_0 (C_2(t) + C_3(t)) = 2 \mu_0 C_2(t)
\end{align}    
arising from the `edge' and `inner' excitation calculated using Eq. \eqref{eq:C(t)_final_pigment}.
Fig. \ref{fig:ISD_spectra} plots the corresponding single-site spectral contribution 
\begin{equation*}
    \mathcal{A}_\mathrm{edge/inner}(\omega) = \textrm{Re}\left[\int_0^\infty C_\mathrm{edge/inner}(t) \mathrm{e}^{-i \omega t} dt\right]
\end{equation*}
for the edge (green line) and inner (blue line) initial conditions. The reconstruction of the total spectrum as the sum of the edge and inner contributions (grey line) agrees with the spectrum arising from the dyadic calculation of the total correlation function (black line) using the four site initial condition (Eq. \eqref{eq:C(t)_psd}). Note that individual spectral contributions (i.e. $\mathcal{A}_\mathrm{edge/inner}(\omega)$) are not themselves proper absorption spectra since initial and final state are not the same.

\subsection{Monte Carlo Sampling Dyadic HOPS}
The total correlation function can be calculated by Monte Carlo sampling over single-site initial conditions. The initial state decomposition of the full correlation function introduces an independent trace over the bath for each initial condition
\begin{equation}
C(t)= \mu_\mathrm{tot} \sum_{\mathbf{d}} A_\mathbf{d} \, \sum_{\mathbf{z}^*} \frac{1}{N_{\mathbf{d},z}} \Bigg[ \frac{\braket{\psi_\mathrm{ex}|\psi_\mathbf{d}(t; \mathbf{z}^*)}}{\frac{1}{2}(\vert \vert  \psi_\mathbf{d}(t; \mathbf{z}^*) \vert \vert^2+1)} \Bigg] e^{i E_{\mathrm{g}} t/\hbar}
\end{equation}
where the sum over noise trajectories ($\mathbf{z}^*$) is independent of the sum over the $N_D$ initial conditions ($\mathbf{d}$). If we assume unbiased sampling with equal sized clusters, then the number of noise trajectories per initial condition is $N_{\mathbf{d},z} = N_{\mathrm{ens}}/N_D$ and the correlation function can be calculated as
\begin{equation}
\label{eq:C(t)_local_montecarlo}
C(t)= \mu_\mathrm{tot}\frac{N_D}{N_{ens}} \sum_{(\mathbf{d}, \mathbf{z}^*)} A_d\, \Bigg[ \frac{\braket{\psi_\mathrm{ex}|\psi_\mathbf{d}(t; \mathbf{z}^*)}}{\frac{1}{2}(\vert \vert  \psi_\mathbf{d}(t; \mathbf{z}^*) \vert \vert^2+1)} \Bigg] e^{i E_{\mathrm{g}} t/\hbar},
\end{equation}
where we Monte Carlo sample $N_{\mathrm{ens}}$ pairs $(\mathbf{d}, \mathbf{z}^*)$ to calculate the full correlation function.

\begin{figure}
\includegraphics{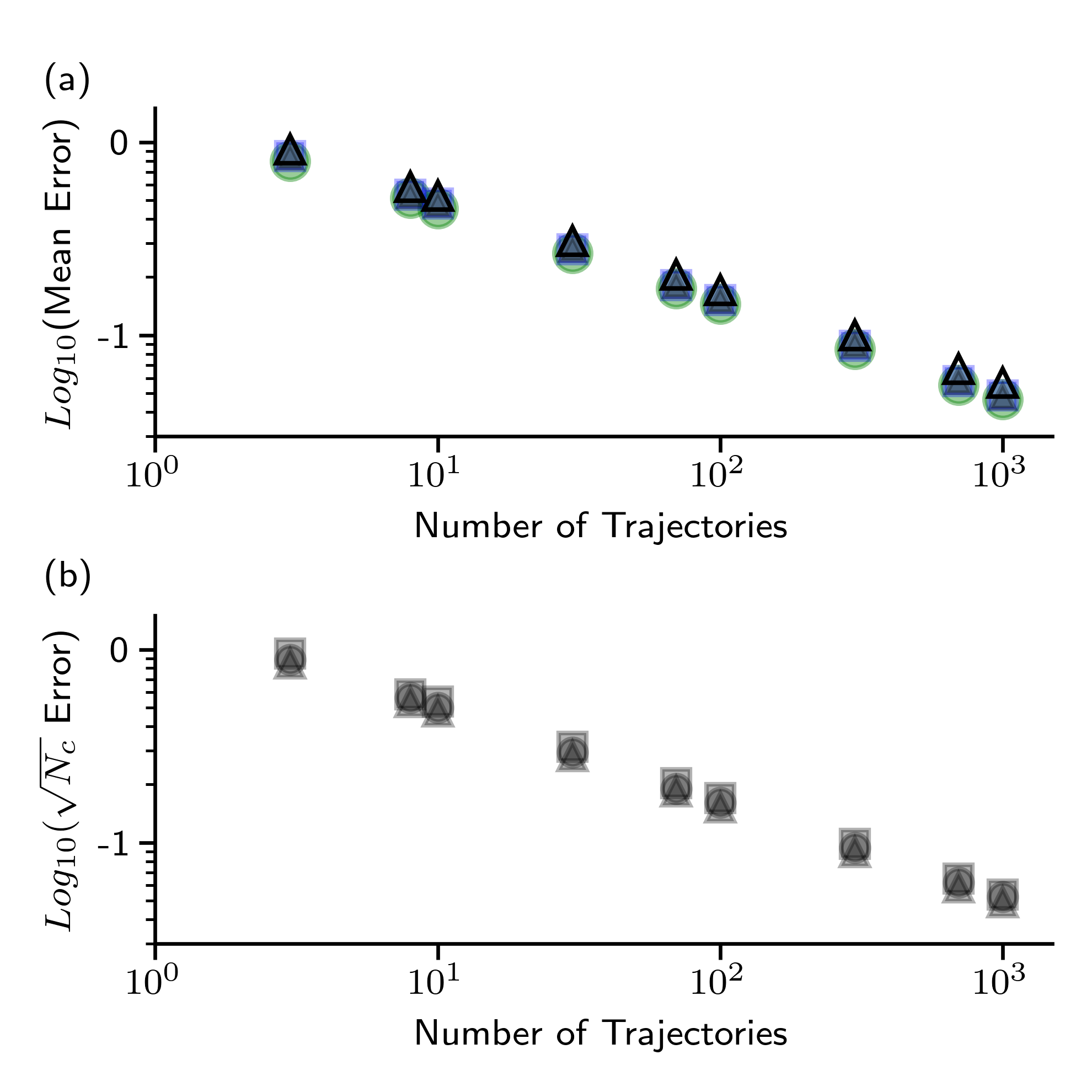}
 \caption{\label{fig:ISD_stat_conv} Statistical convergence of Monte Carlo sampling over the initial state decomposition. (a) Bootstrapped spectral error for the total absorption spectrum calculated with single site initial conditions (black triangles, $N_c=1$) either without (filled) or with (open) static disorder. Similar statistical convergence is observed for the edge (green circles) and inner (blue squares) single site initial conditions. (b) Comparison of the statistical error for the total spectrum normalized by the square-root of the number of pigments in each cluster ($\sqrt{N_c}$) when using single site ($N_c=1$, triangles), pair ($N_c=2$, circles), and all four site ($N_c=4$, squares) initial conditions. Parameters: $V=-100 \textrm{ cm}^{-1}$, $\lambda = 35 \textrm{ cm}^{-1}$, $\Gamma=50 \textrm{ cm}^{-1}$, $\nu = 10 \textrm{ cm}^{-1}$, $T=295 \textrm{ K}$, and $k_{\mathrm{max}}=5$.}   
 \end{figure}
\paragraph{Example:}
Here, we perform calculations using the same 4 site linear chain considered above. Fig.~\ref{fig:ISD_stat_conv}a compares the bootstrapped spectral error for the total spectrum using Eq. \eqref{eq:C(t)_local_montecarlo} with single site initial conditions (black filled triangles)
\begin{equation}
\label{eq:diff_measure}
    \mathrm{error} = \frac{1}{\int_{\omega_\mathrm{min}}^{\omega_\mathrm{max}} \mathcal{A}_{\mathrm{ref}}(\omega)\mathrm{d}\omega}\int_{\omega_\mathrm{min}}^{\omega_\mathrm{max}} \big|\mathcal{A}(\omega)-\mathcal{A}_{\mathrm{ref}}(\omega)\big|\mathrm{d}\omega
\end{equation}
as a function of the number of trajectories sampled from the combined $(n, \mathbf{z}^*)$ space. We find that the statistical convergence of the total correlation function is equivalent to that of the edge (green circles) and inner (blue squares) spectral components calculated independently using Eq. \eqref{eq:C(t)_final_pigment}.

One advantage of Monte Carlo sampling is the trivial incorporation of an additional sampling over disorder in the system Hamiltonian ($\hat{H}_{s}$). Fig.~\ref{fig:ISD_stat_conv}a shows the statistical convergence of the total spectrum calculated with single-site initial conditions in the presence of Gaussian distributed disorder on site energies (black open triangles) is almost unchanged compared to the homogeneous case (black filled triangles). For the disordered calculations, site energies form a Gaussian distribution with mean value of zero and a standard deviation (SD) of 100 cm$^{-1}$.

Finally, we find that the statistical error of the Monte Carlo sample for a given number of trajectories is inversely proportional to the size of the clusters ($N_c = N_\mathrm{pig}/N_D$) used for the initial state decomposition. In Fig.~\ref{fig:ISD_stat_conv}b, we compare the normalized statistical error ($\sqrt{N_c} \cdot$ Error) as a function of the number of independent trajectories ($N_{\mathrm{traj}}$) when using different initial conditions: single site  (triangles, $N_c =1$), pairs (circles, $N_c = 2$), and all four sites (squares, $N_c = 4$). 
The equivalence of these three results shows that increasing the size of the clusters decreases the number of trajectories required to reach a given statistical error. 
Thus, when using an equation of motion where the delocalization extent of the initial state does not change the computational expense, the initial state decomposition provides no advantage.

\subsection{Monte Carlo Sampling Dyadic Adaptive HOPS}
\begin{figure}
\includegraphics{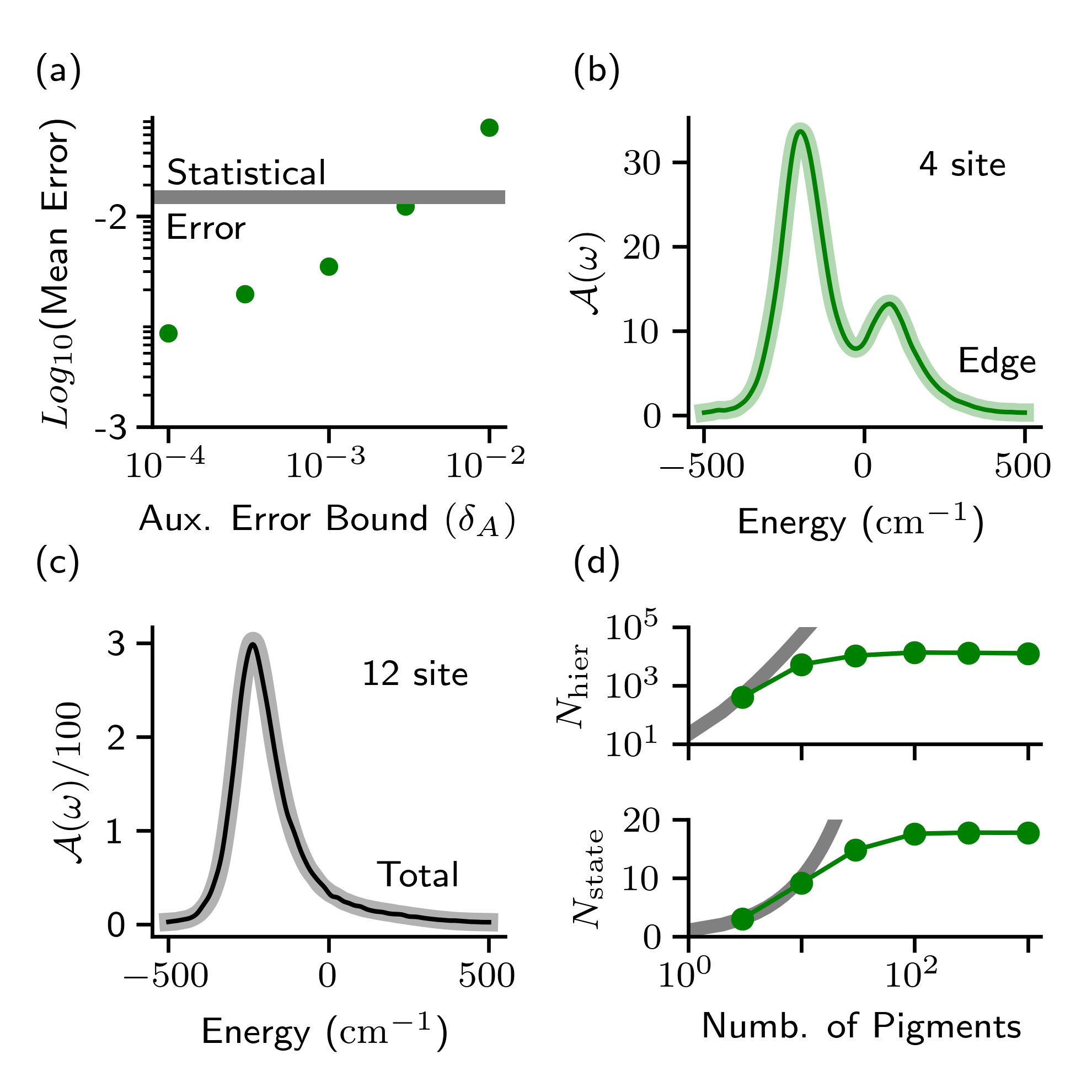}
\caption{\label{fig3} Comparing dyadic HOPS and DadHOPS for N-site linear chains. (a) Mean error of the edge spectrum with respect to the auxiliary error bound ($\delta_A$) for a 4-site chain. The solid grey line represents the statistical error. (b) The edge spectral component for a 4-site chain calculated with converged DadHOPS parameters (thin line) and corresponding dyadic HOPS calculation (thick transparent line). (c) Total absorption spectrum for a 12-site chain calculated using DadHOPS with (thin line) and without state adaptivity (thick line). (d) Average number of auxiliary wave functions (top) and state basis elements (bottom) for linear chains of different lengths required for calculations using DadHOPS (green) compared to dyadic HOPS (thick grey). Parameters: $V=-100 \textrm{ cm}^{-1}$, $\lambda = 35 \textrm{ cm}^{-1}$, $\Gamma=50 \textrm{ cm}^{-1}$, $\nu = 10 \textrm{ cm}^{-1}$, $T=295 \textrm{ K}$, and $k_{\mathrm{max}}=5$.  }
\end{figure}

The local correlation functions arising from the initial state decomposition can be efficiently simulated using DadHOPS. 
Fig.~\ref{fig3}a shows how, for a 4-site linear chain, the error of the edge component decreases with the derivative error bound $\delta_A$. 
Here, we consider DadHOPS calculations converged when the adaptive error is lower than the statistical error for the associated ensemble of $10^4$ trajectories (Fig.~\ref{fig3}a, grey horizontal line). 
In Fig.~\ref{fig3}b the edge spectral contribution calculated with converged DadHOPS parameters (thin line, $\delta_A = 10^{-3}$) reproduces the corresponding dyadic HOPS calculation (thick transparent line). 

The DadHOPS algorithm combined with the initial state decomposition allows us to calculate much larger system sizes then would be possible with dyadic HOPS directly. 
As a first demonstration of scaling we consider a 12-site linear chain where the basis for the full HOPS calculation has more than $10^6$ elements. 
Fig.~\ref{fig3}c plots the total absorption spectrum calculated with DadHOPS (thick line, $\delta_A = 10^{-3}, \,\, \delta_S=0$) using the full initial condition ($\ket{\psi_0} = \frac{1}{\sqrt{12}}\sum_{n=1}^{12} \ket{n}$). 
We can reproduce this spectrum by Monte Carlo sampling over single-site initial conditions and incorporating adaptivity in both the auxiliary and state basis (Fig.~\ref{fig3}c, thin line, $\delta_A = \delta_S = 10^{-3}$). 
Combining adaptivity in both the auxiliary and state basis with localized initial states can achieve size-invariant scaling for sufficiently large aggregates. 
Fig.~\ref{fig3}d plots the number of state basis elements (bottom) and the number auxiliary wave functions (top) required for both the full dyadic HOPS (thick grey lines) and the DadHOPS algorithm (thin green lines) as a function of the number of sites in the linear chain. 

Here, we have demonstrated that the advantage of the localized wave functions introduced by the initial state decomposition is realized by combining it with the DadHOPS equation-of-motion that leverages locality to reduce computational expense.

\section{Application} \label{Sec:Application}
\subsection{Photosystem I (PSI)}
\begin{figure}
\includegraphics{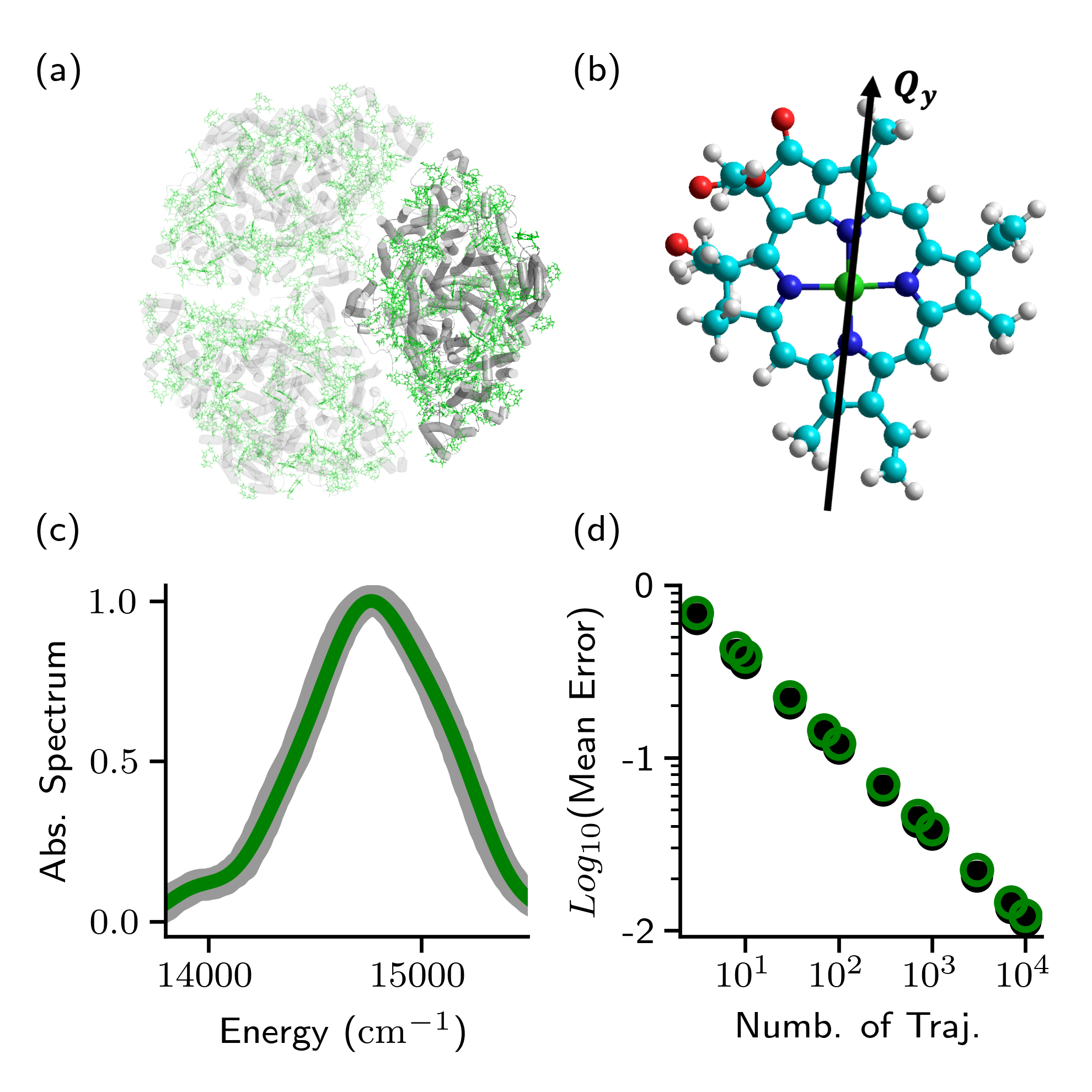}
\caption{\label{fig:PSI} Simulating linear absorption for Photosystem I (PSI).   (a) A PSI core complex trimer is shown with a single monomer highlighted; the simulations are for one monomer containing 96 pigments, using cyclic boundary conditions to represent the trimer. (b) A chlorophyll \emph{a} molecule with the phytyl tail truncated for clarity is show along with the Qy transition dipole vector. (c) Comparison of absorption spectrum simulated using HEOM (grey)\cite{KRAMER2018} and Dyadic HOPS (green). (d) Bootstrapped spectral error for randomly assigning pigments to clusters (black solid circles) and using clusters defined by strong coupling (green open circles). Parameters: $\lambda = 35 $ cm$^{-1}$, $\Gamma = 50$ cm$^{-1}$, $T=300$ K, $\Gamma_\mathrm{mark} = 500$ cm$^{-1}$, SD$=150$ cm$^{-1}$, and $k_{\mathrm{max}}=1$.}
\end{figure}

Photosystem I (PSI) is a pigment-protein complex containing 96 chlorophyll per monomer and usually found as a trimer in higher plants and algae (Fig \ref{fig:PSI}a).\cite{Jordan2001}  
The complicated spatial arrangement of chlorophyll make this an ideal test system for exploring the behavior of the initial state decomposition with respect to different definitions of clusters. 
Here, we use a previously reported Hamiltonian\cite{Yin2007} where chlorophyll excitation energies ($E_n$) and excitonic couplings ($V_{n,m}$) were calculated using time-dependent density function theory (TDDFT) evaluated by Gaussian16 \cite{g16} with the CAM-B3LYP density functional\cite{Yanai2004} and the 6-31G$^{*}$ basis set;\cite{Hehre1972} this approach is known to provide quality descriptions of both vibrational and vibronic processes apparent from the high-resolution absorption and emission spectra of chlorophyll-type molecules.\cite{Reimers2022} The Hamiltonian is provided in the Supporting Information. The system-bath coupling is described by a Drude-Lorentz spectral density
\begin{equation}
J_n(\omega) = \frac{2 \lambda \Gamma \omega}{\omega^2  + \Gamma^2}
\end{equation}
and the corresponding bath correlation function is given in a high-temperature approximation as
\begin{equation}
    \alpha_n(t) = (2 \lambda / \beta - i \lambda \Gamma) e^{-\Gamma t/\hbar} + i \lambda \Gamma e^{-\Gamma_\mathrm{mark} t/\hbar}
\end{equation}
where the second exponential is included to ensure the imaginary component of $\alpha_n(t)$ is continuous at time $t=0$. To agree with previous HEOM calculations,\cite{KRAMER2018} we introduce a Gaussian distributed static disorder on the chlorophyll site energies with a standard deviation (SD) of 150 cm$^{-1}$, we approximate the true chlorophyll Qy transition dipole moment (Fig. \ref{fig:PSI}b) by the vector defined by the position of the NA and NC nitrogen atoms (IUPAC notation for chlorophyll \emph{a}),\cite{blankenship2021molecular} and we set the spectral density parameters to be:  $\lambda = 35 $ cm$^{-1}$, $\Gamma = 50$ cm$^{-1}$, $T=300$ K, and $\Gamma_\mathrm{mark} = 500$ cm$^{-1}$. 

The varied alignments of the chlorophyll Qy dipole moments throughout PSI introduces an orientation dependence into the linear absorption spectrum. For an isotropic distribution of PSI complexes, we calculate the linear absorption spectrum from the average of the dipole autocorrelation function for the x, y, and z polarized electric fields ($\bfepsilon$)
\begin{equation}
C^{(\bfepsilon)}(t)= \mu_\mathrm{tot, \bfepsilon} \sum_{d=1}^{N_D}  \, \left(\frac{A_{d,\bfepsilon}}{N_{d,z}} \right) \, \sum_{\mathbf{z}^*} \Bigg[ \frac{\braket{\psi_\mathrm{ex}|\psi_d(t; \mathbf{z}^*)}}{\frac{1}{2}(\vert \vert  \psi_d(t; \mathbf{z}^*) \vert \vert^2+1)} \Bigg] e^{i E_{\mathrm{g}} t/\hbar}.
\end{equation}
Fig. \ref{fig:PSI}c shows that, as expected, the HOPS calculations (green line) reproduce previous HEOM calculations from Ref.~\onlinecite{KRAMER2018} (grey line). 

To what extent does the specific choice of clusters matter for a given cluster size? In the case of a linear chain, the choice of clusters to be sets of adjacent pigments may appear obvious. There is no equivalently obvious choice for the heterogeneously coupled pigments in PSI. Fig. \ref{fig:PSI}d compares the statistical convergence of the linear absorption spectrum for two different definitions of clusters: the first randomly assigns four pigments to each cluster (black solid circles), while the second constructs clusters of 4 strongly interacting pigments (green open circles) using an algorithm described in Appendix \ref{app:ClusterDefinition}. There is no appreciable difference in the convergence between randomly assigning pigments to clusters or using clusters defined by strong coupling. Combined with the results presented in Fig. \ref{fig:ISD_stat_conv}b showing square-root-scaling of error with cluster size, our calculations suggest that using a cluster initial condition only acts to increase the effective number of combined $(n, \mathbf{z}^*)$ pairs that are being sampled.

\subsection{Perylene bis-imide (PBI)}
Perylene bis-imide (PBI) is a family of pigments that can form both J- and H-type aggregates in solution.\cite{Dehm2007,Chen2007Chem,Marciniak2010JPCA,Schrter2015} J-aggregates of PBI-1 have been the object of particular study and show a strong vibronic progression in their linear absorption sprectra. Previous time-dependent density functional theory (TDDFT) calculations of PBI-1 have characterized electronic coupling between adjacent monomers ($V \approx -500 \textrm{ cm}^{-1}$) and 28 intramolecular vibrational modes with appreciable Huang-Rhys factors.\cite{Ambrosek2012} This vibrational structure has served as a starting point for different calculations of linear absorption spectra for PBI-1 aggregates using techniques such as multiconfiguration time-dependent Hartree (MCTDH), \cite{Ambrosek2012} multilayer MCTDH (ML-MCTDH),\cite{Seibt2009} and time-dependent density matrix renormalization group theory (TD-DMRG).\cite{Ren2018} Calculations of the exciton dynamics in aggregates ranging in size from two to 25 monomer units have also been reported with varying degrees of complexity in their description of the molecular vibrations. \cite{Kundu2021,Kundu2021PCCP,Schrter2015} 

We have constructed a minimal bath correlation function composed of two modes to describe the vibrational environment of PBI. We include a vibration with a central frequency near 1500 cm$^{-1}$ to account for the group of strongly-coupled vibrations ranging from 1370-1630 cm$^{-1}$ previously reported from TDDFT calculations.\cite{Ambrosek2012}  We also include a low-frequency vibration to account for the broad dissipative environment. The final parameters (Table \ref{tab:2-exp_diss}) were selected to agree with the available experimental data. 
\begin{center}
\begin{table}
\begin{tabular}{||c|c|c||} 
 \hline
 Mode & g (cm$^{-2}$) & $\gamma$ (cm$^{-1}$) \\ [1ex] 
 \hline\hline
1 & \hspace{0.5 cm} $1.2\times10^5$ \hspace{0.5 cm} & \hspace{0.5 cm} $50 + 170i$ \hspace{0.5 cm} \\
  \hline
2 & \hspace{0.5 cm} $1.6\times 10^6$ \hspace{0.5 cm} & \hspace{0.5 cm} $100 + 1550i$ \hspace{0.5 cm}\\ 
 \hline
\end{tabular}
 \caption{\label{tab:2-exp_diss} The exponential parameters describing the vibrational correlation function of PBI.}
 \end{table}
\end{center}

\begin{figure}
\includegraphics{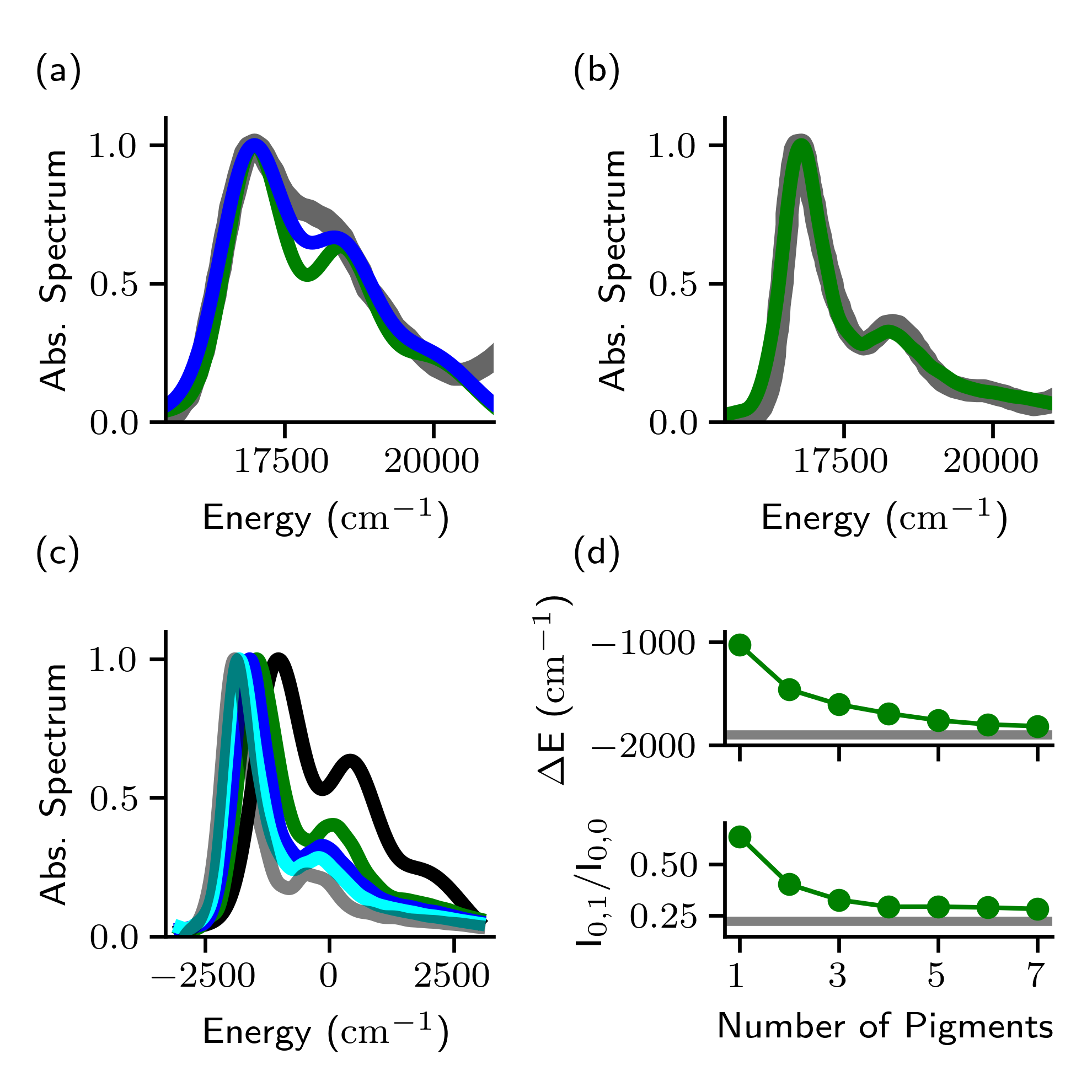}
\caption{\label{fig:PBI} Simulating linear absorption for perylene bis-imide (PBI) J-aggregates. (a) Compares the experimental absorption spectra from dilute solution (0.0016 mM, grey)]\cite{Marciniak2010JPCA} with calculations broadened by either $\mathrm{SD}= 300$ cm$^{-1}$ (green) or $400$ cm$^{-1}$ (blue). (b) Compares the experimental absorption spectra from concentrated solution (0.16 mM, grey)\cite{Marciniak2010JPCA} with a calculated trimer spectra (green). (c) Compares the lineshape calculated for aggregates containing one (black), two (green), three (blue), seven (cyan), and 1000 pigments (grey). (d) Plots the change in the 0-0 transition peak shift ($\Delta E$, top) and the ratio of the 0-1 peak intensity to the 0-0 peak intensity ($I_{0,1}/I_{0,0}$, bottom) as a function of the number of molecules in the aggregate ($N_\mathrm{pig}$). Parameters: $k_{\mathrm{max}} = 6$.
}
\end{figure}

Our minimal description of the system-bath correlation function is capable of reproducing the major features of both the experimentally measured monomer PBI spectrum (Fig. \ref{fig:PBI}a) and the aggregate spectrum (Fig. \ref{fig:PBI}b). The monomer spectrum shows a broad lineshape and is better reproduced when the  Gaussian disorder of the vertical excitation energies has a standard deviation of $\mathrm{SD}= 400 \textrm{ cm}^{-1}$ (blue line) compared with the $\mathrm{SD} = 300 \textrm{ cm}^{-1}$ (green line) we use for the aggregate spectrum. We can further improve the agreement between the simulated and experimental monomer spectra by introducing additional moderate frequency modes (data not shown), but this additional complexity was not required for the reproduction of the experimental aggregate spectrum. We smooth the aggregate spectra reported here by weighting the time-domain dipole auto-correlation function with a cosine appodization window\cite{hoch1996nmr}
\begin{equation}
    \Theta(t) = 
    \begin{cases}
      \cos(\frac{\pi}{2}t/t_{\mathrm{max}}) &  t\leq t_{\mathrm{max}} \\
      0        & t>t_{\mathrm{max}}
    \end{cases}
\end{equation}
that goes to zero at the last time point of the calculated trajectory ($t_{\mathrm{max}}$). The use of this window function suppresses noise in the calculated spectra that arises from the combination of zero-padding\cite{hoch1996nmr} and the incomplete cancellation of the correlation function at long times due to finite sampling of the trajectory ensembles.

The advantage of the adaptive dyadic HOPS formalism is the ability to calculate even very large molecular aggregates efficiently. The size of a molecular aggregate is often important for capturing the influence of exciton delocalization on both the 0-0 peak position and the relative magnitude of vibronic side-bands.\cite{Ghosh2020} Fig. \ref{fig:PBI}c compares the monomer (black line), dimer (green line), trimer (blue line), and heptamer (cyan line) lineshapes with that calculated for a 1000-site linear chain (grey line). The magnitude of the vibronic side-band decreases rapidly from the monomer to trimer and then changes slow with increasing chain length. A similar trend is seen for the red-shift of the 0-0 peak. Fig. \ref{fig:PBI}d quantifies this effect by plotting the central position of the 0-0 peak ($\Delta E$, top) and the ratio of the intensity of the 0-1 peak to the intensity of the 0-0 peak ($I_{0,1}/I_{0,0}$, bottom) as a function of the number of pigments in the aggregate. In both cases, the grey lines represents the asymptotic limit of a 1000-site linear chain. 

\begin{figure}
\includegraphics{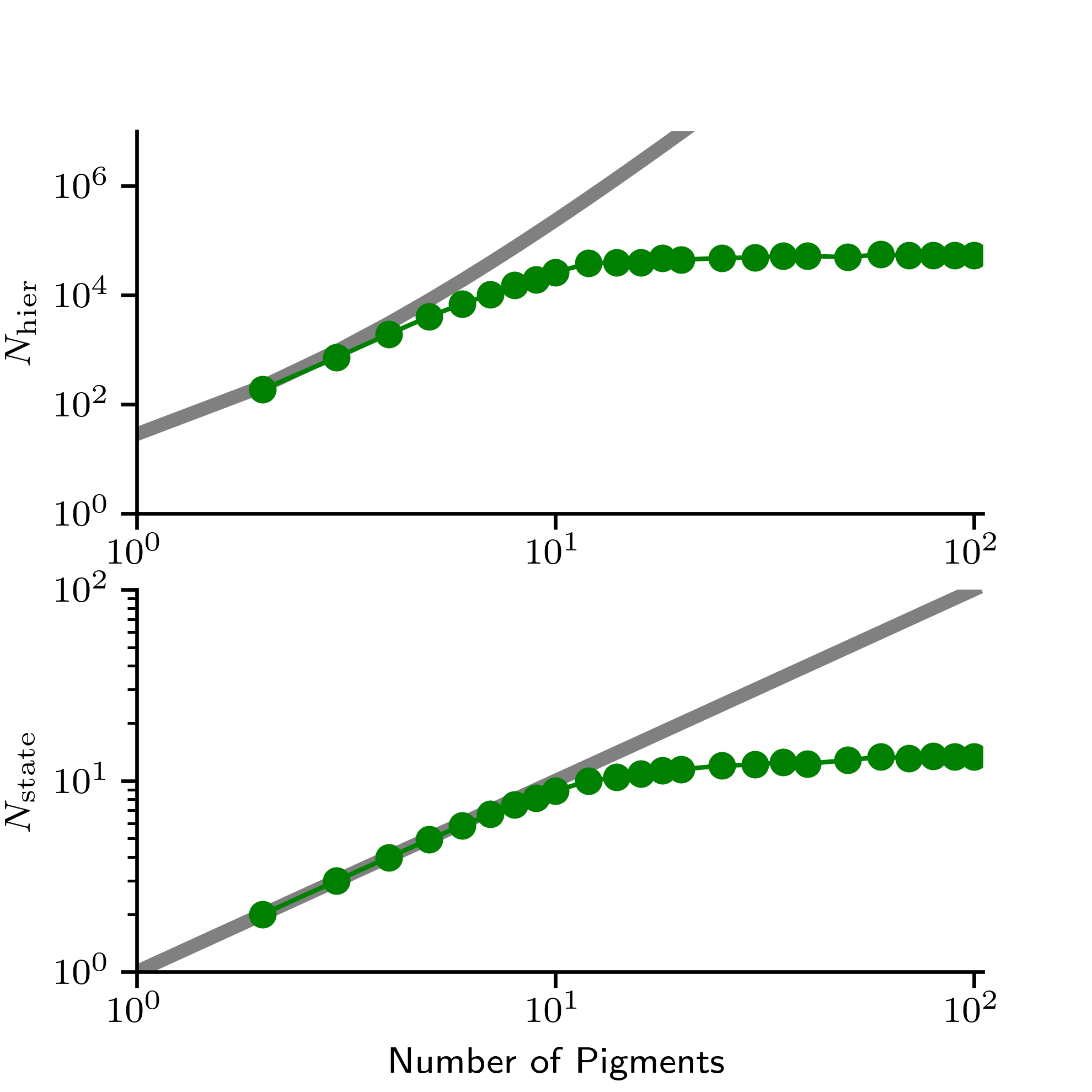}
\caption{\label{Fig:PBI-Nhier-Nstate} Scaling of DadHOPS basis size for PBI aggregate. Average number of auxiliary wave functions (top) and states (bottom) for different lengths of PBI chains required for calculations using DadHOPS (green) compared to dyadic HOPS (grey). }
\end{figure}

Finally, let us consider the relationship between aggregate size, locality, and the onset of size-invariant scaling in the DadHOPS equation-of-motion. When the adaptive basis size stops increasing with the length of the aggregate, the extent of exciton delocalization is sufficiently small that most trajectories will never sample the edge of the aggregate. This implies that for aggregates sufficiently large that the basis is size-invariant the spectroscopic signatures should also be size-invariant. Fig. \ref{Fig:PBI-Nhier-Nstate} shows that by $N_\mathrm{pig}=10$ the average size of the adaptive site basis begins to plateau before reaching a size-invariant value of 13 for chains of 100 PBI molecules. 
This is consistent with the convergence of the spectral features which are close to the asymptotic values at $N_\mathrm{pig}=7$ (Fig.~\ref{fig:PBI}d) and the expectation that complete size-invariance will occur when only a relatively small set of trajectories will sample pigments on the edge of the chain. We suggest, then, that DadHOPS can be thought of as introducing a dynamic separation of length-scales where the basis size required to capture the spectral features is solved on-the-fly rather than being asserted \emph{a priori}. As a result, DadHOPS is advantageous for simulating realistic molecular aggregates, particularly in the presence of structural disorder, where separations of length scales can be obscure.        

\section{Conclusions} \label{Sec:Conclusions}
Here we have presented a new algorithm that combines the dyadic adaptive HOPS (DadHOPS) equations with a initial state decomposition and is capable of calculating optical linear absorption spectra for mesoscale molecular aggregates.
Our approach introduce an dynamic separation of length-scales by adaptively constructing a reduced basis to describe the time-dependence of local contributions to the dipole autocorrelation function. 
The adaptive basis construction is capable of achieving size-invariant scaling (i.e. $\mathcal{O}(1)$) with the number of molecules for sufficiently large aggregates.  
We have applied the initial state decomposition to the 96 chlorophyll photosystem I core complex and found that the specific choice of pigment clusters is not essential to the statistical convergence of the calculations. 
We calculated 1000 molecule J-aggregate of perylene bis-imide (PBI) with DadHOPS and characterized how the absorption spectra changes with aggregate size. 
Because locality cannot reduce the computational expense of describing individual pigments, DadHOPS remains limited to calculations where each thermal reservoir has a small number of exponential modes. 
In the future, it may be possible to describe more complex thermal environment by incorporating techniques such as tensor contraction.\cite{gaoNonMarkovianStochasticSchrodinger2022}

\begin{acknowledgments}
AE acknowledges support from the DFG via a Heisenberg fellowship (Grant No EI 872/10-1).
TG, EJT, and DIGB acknowledge support from the Robert A.~Welch Foundation (Grant N-2026-20200401) and start-up funds from Southern Methodist University. DIGB acknowledges  the US National Science Foundation CAREER Award under Grant No.~CHE-2145358.
\end{acknowledgments}

\section*{Data Availability Statement}
The data that support the findings of this study are available from the corresponding author upon reasonable request.

\appendix
\section{Dyadic HOPS with Generalized Initial State Decomposition}
The dipole autocorrelation function 
\begin{eqnarray}
C(t)
=\mathrm{Tr}\left\lbrace\hat{\mu}_{\mathrm{eff}}\,{e}^{-i\hat{H}t/\hbar}\big(\hat{\mu}_\mathrm{eff}{|}\mathrm{g}\rangle\langle{\mathrm{g}}|\otimes\hat{\rho}_{\mathrm{B}}\big) e^{i\hat{H}t/\hbar}\right\rbrace
\end{eqnarray}
depends on the time-evolution of the first-order density matrix  $\hat{\mu}_{\textrm{eff}}\ket{g}\bra{g}$ arising from one interaction with the electric field. The Non-Markovian Quantum State Diffusion (NMQSD) formalism that gives rise to the HOPS equation, however, can only time-evolve pure state density matrices. The pure state decomposition introduced in Ref.~\onlinecite{Hartmann2021} rewrites the initial first-order density matrix into a sum of pure state density matrices which can then be treated within the NMQSD (or HOPS) formalism. 
Here, we use the pure state decomposition to rewrite the dipole autocorrelation function into a sum of locally excited correlation functions that can be efficiently simulated using the DadHOPS equation-of-motion. 

We begin by decomposing the collective transition dipole operator 
\begin{equation}
\hat{\mu}_\mathrm{eff}=\sum_n^N (\bfmu_n\cdot\bfepsilon) \,|n\rangle\langle{\mathrm{g}}|+ h.c.,
\end{equation}
into an arbitrary (finite) set of interaction operators ($\hat{\sigma}_a$) and real-valued weights ($A_a$) defined to ensure that 
\begin{equation}
\hat{\mu}_{\mathrm{eff}} = \sum_a A_a \hat{\sigma}_a
\end{equation}
and  $\ket{\psi_{a}} = \hat{\sigma}_a\ket{g}$ is a normalized wave function. 
The dipole autocorrelation function can then be rewritten as 
\begin{equation}
C(t) = \sum_a A_a C_a(t)
\end{equation}
where
\begin{equation}
    C_a(t) = \mathrm{Tr}\left\lbrace\hat{\mu}_{\mathrm{eff}}\,{e}^{-i\hat{H}t/\hbar}\big( \hat{\sigma}_\mathrm{a}{|}\mathrm{g}\rangle\langle{\mathrm{g}}|\otimes\hat{\rho}_{\mathrm{B}}\big) e^{i\hat{H}t/\hbar}\right\rbrace.
\end{equation}
The decomposition expressed here is generic and does not require - for example - that the interaction operators be orthogonal. We will proceed in our derivation with these generic interaction operators and then introduce the specific expressions for two convenient special cases at the end. 

We can then define a set of pure states 
\begin{equation}
\label{eq:pure_state_m}
    \ket{v_{\eta,a}} = \frac{1}{\sqrt{2}}\Big(\eta\ket{\mathrm{g}} + 
\ket{\psi_a}\Big)
\end{equation}
which can be used to reconstruct the initial first-order density matrix associated with the correlation function for each excitation operator
\begin{equation}
\label{eq:pure_decomp}
\rho_a(t=0) = \hat{\sigma}_\mathrm{a} \ket{\mathrm{g}}\bra{\mathrm{g}}=
 \sum_{\eta\in \{\pm 1, \pm i\}} \frac{\eta}{2} \ \ket{v_{\eta, a}}\bra{v_{\eta,a }}.
\end{equation}
To prove the second equality, we expand the summand 
\begin{equation}
\begin{aligned}
 \frac{\eta}{2} \ \ket{v_{\eta, a}}\bra{v_{\eta,a }} =   \frac{1}{4} \Big(&\eta \eta \eta^* \ket{g}\bra{g} + \eta \eta\ket{g}\bra{\psi_a} \\
&+ \eta \eta^*\ket{\psi_a}\bra{g} + \eta \ket{\psi_a}\bra{\psi_a}\Big) 
\end{aligned}
\end{equation}
and note that 
\begin{eqnarray*}
    \sum_{\eta\in \{\pm 1, \pm i\}} \eta \eta \eta^* = 0, \sum_{\eta\in \{\pm 1, \pm i\}} \eta \eta = 0 \\
    \sum_{\eta\in \{\pm 1, \pm i\}} \eta \eta^* = 4, \sum_{\eta\in \{\pm 1, \pm i\}} \eta  = 0
\end{eqnarray*}
which reduces the right-hand side of Eq. \eqref{eq:pure_decomp} to 
\begin{equation}
    \sum_{\eta\in \{\pm 1, \pm i\}} \frac{\eta}{2} \ \ket{v_{\eta, a}}\bra{v_{\eta,a }} = \ket{\psi_a}\bra{g} = \hat{\sigma}_a \ket{g}\bra{g}.
\end{equation}

We can now proceed to write down the standard dyadic HOPS expressions for the individual contributions to the dipole autocorrelation function. 
We begin by rewriting the components of the dipole autocorrelation function as
\begin{equation}
    C_a(t) = \sum_{\eta\in \{\pm 1, \pm i\}}   \mathrm{Tr}\left\lbrace\hat{\mu}_{\mathrm{eff}}\,{e}^{-i\hat{H}t/\hbar}\big( \frac{\eta}{2} \ket{v_{\eta, a}}\bra{v_{\eta,a }}\otimes\hat{\rho}_{\mathrm{B}}\big) e^{i\hat{H}t/\hbar}\right\rbrace
\end{equation}
which is equal to 
\begin{equation}
\label{eq:Ct_eta_a_hops}
    C_a(t) = \sum_{\eta\in \{\pm 1, \pm i\}}   \mathbb{E}\Bigg[\frac{\frac{\eta}{2}\braket{v_{\eta, a}(t, \mathbf{z}^*)|\hat{\mu}_{\mathrm{eff}}|v_{\eta, a}(t, \mathbf{z}^*)}}{\braket{v_{\eta, a}(t, \mathbf{z}^*)|v_{\eta, a}(t, \mathbf{z}^*)}}\Bigg]
\end{equation}
where $\ket{v_{\eta, a}}$ is time-evolved according to the non-linear HOPS equation (Eq.~\eqref{eq:NormNonLinearHops}). As written, the time-evolution of the pure state $\ket{v_{\eta, a}}$ is performed using a system Hamiltonian ($\hat{H}_S$) that contains both the ground and excited electronic states. However, since the system Hamiltonian does not couple the electronic ground and excited states, the time evolution of the two components of $\ket{v_{\eta, a}}$ can be decoupled giving
\begin{equation}
\label{eq:pure_state_a}
    \ket{v_{\eta,a}(t;\mathbf{z}^*)} = \frac{1}{\sqrt{2}}\Big(\eta\ket{\mathrm{g}}e^{-iE_g t/\hbar} + 
\ket{\psi_a(t;\mathbf{z}^*)}\Big)
\end{equation}
where $\ket{\psi_a(t;\mathbf{z}^*)}$ is propagated using the non-linear HOPS equation where $\hat{H}_S$ includes only the first excitation manifold and the expectation value of the system-bath coupling operator in Eq.~(\ref{eq: dyadicHOPS_renormalization}) is redefined as 
\begin{equation}
    \langle\hat{L}^{\dagger}_{n}\rangle_{t} = \frac{\langle \psi^{(\vec{0})}(t; \mathbf{z}^*) \vert \hat{L}^{\dagger}_{n}\vert \psi^{(\vec{0})}(t; \mathbf{z}^*) \rangle}{\braket{\psi^{(\vec{0})}(t; \mathbf{z}^*) \vert \psi^{(\vec{0})}(t; \mathbf{z}^*)}+1}
\end{equation}
where the change in the denominator arises from the presence of the ground-state component of the wave function. 

We can use Eq.~\eqref{eq:pure_state_a} and the definition of the collective dipole operator ($\hat{\mu}_{\mathrm{eff}}\ket{g} = \mu_{\mathrm{tot}}\ket{\psi_{\mathrm{ex}}}$, Eq.~\eqref{eq:collective_Transition_Operator}) to rewrite the numerator of Eq.~\eqref{eq:Ct_eta_a_hops} as
\begin{eqnarray*}
    \frac{\eta}{2} \braket{v_{\eta, a}(t; \mathbf{z}^*)|\hat{\mu}_{\mathrm{eff}}|v_{\eta, a}(t; \mathbf{z}^*)} = & \mu_{\mathrm{tot}}\frac{\eta\eta}{4}\braket{\psi_a(t;\mathbf{z}^*) | \psi_{\mathrm{ex}}}e^{-iE_g t/\hbar} 
    \\
    &+ \mu_{\mathrm{tot}}\frac{\vert \eta \vert^2}{4} \braket{\psi_{\mathrm{ex}} | \psi_a(t;\mathbf{z}^*)} e^{iE_g t/\hbar}
\end{eqnarray*}
and the denominator as
\begin{equation}
    \braket{v_{\eta, a}(t; \mathbf{z}^*)|v_{\eta, a}(t; \mathbf{z}^*)} = \frac{1}{2}\big(1 + \vert \vert \psi_a(t; \mathbf{z}^*) \vert \vert^2 \big).
\end{equation}
Noting that the denominator does not depend on $\eta$, the cancellation in the numerator due to summation over $\eta\in \{\pm 1, \pm i\}$ gives
\begin{equation}
\label{eq:C(t)_final_normalized_a}
C_a(t)=   \mu_{\mathrm{tot}}\mathbb{E}\Big[\frac{\braket{\psi_\mathrm{ex} \vert \psi_a(t;\mathbf{z}^*)}}{\frac{1}{2}\left(||\psi_a(t,\mathbf{z}^*)||^2 + 1\right)}\Big]e^{i E_{\mathrm{g}} t/\hbar}.
\end{equation}
This equation is equivalent to the standard dyadic HOPS expression, but now propagating a component of the total correlation function defined by a decomposition of the collective transition dipole operator into a sum of excitation operators. 

\subsubsection{Cluster Decomposition}
One simple decomposition of the effective collective transition dipole operator is to expand in clusters of pigments
\begin{align}
        \hat{\mu}_{\mathrm{eff}} =& \sum_n^N (\bfmu_n\cdot\bfepsilon) \,|n\rangle\langle{\mathrm{g}}|+ h.c. 
        \\
        =& \sum_{\mathbf{d}} A_{\mathbf{d}} \sum_{d\in\mathbf{d}} \frac{\bfmu_{d}\cdot\bfepsilon}{A_{\mathbf{d}}} (|d\rangle\langle{\mathrm{g}}|+ |{\mathrm{g}}\rangle\langle d|)
        \\
        =& \sum_{\mathbf{d}} A_{\mathbf{d}} \hat{\sigma}_{\mathbf{d}} 
\end{align}
where
\begin{equation}
    \hat{\sigma}_d = \sum_{d\in\mathbf{d}} \frac{\mathbf{\mu_{d}}\cdot\bfepsilon}{A_{\mathbf{d}}} (|d\rangle\langle{\mathrm{g}}|+ h.c.)
\end{equation}
and 
\begin{equation}
    A_{\mathbf{d}} =  \sqrt{\sum_{d\in\mathbf{d}} (\bfmu_{d}\cdot\bfepsilon)^2}.
\end{equation} 
The total correlation function can then be decomposed as 
\begin{equation}
\label{eq:C(t)_final_cluster}
C(t)= \mu_{\mathrm{tot}} \sum_{\mathbf{d}} A_{\mathbf{d}}  \mathbb{E}\Big[\frac{\braket{\psi_\mathrm{ex} \vert \psi_{\mathbf{d}}(t;\mathbf{z}^*)}}{\frac{1}{2}\left(||\psi_{\mathbf{d}}(t;\mathbf{z}^*)||^2 + 1\right)}\Big]e^{i E_{\mathrm{g}} t/\hbar}
\end{equation}
where $\ket{\psi_{\mathbf{d}}(t;\mathbf{z}^*)}$ is the state $\ket{\psi_{\mathbf{d}} }= \hat{\sigma}_{\mathbf{d}} \ket{g}$ time-evolved according to the dyadic HOPS equation. 

\subsubsection{Pigment Decomposition}
If each cluster is composed of a single pigment, then Eq.~\eqref{eq:C(t)_final_cluster} simplifies to
\begin{equation}
C(t)=\mu_{\mathrm{tot}} \sum_n (\bfmu_n\cdot\bfepsilon)  \mathbb{E}\Big[\frac{\braket{\psi_\mathrm{ex} \vert n(t;\mathbf{z}^*)}}{\frac{1}{2}\left(||n(t;\mathbf{z}^*)||^2 + 1\right)}\Big]e^{i E_{\mathrm{g}} t/\hbar}
\end{equation}
where $\ket{n(t;\mathbf{z}^*)}$ is the single-pigment excitation state $\ket{n}$ time-evolved according to the dyadic HOPS equation.

\section{Defining Strongly Interacting Clusters}
\label{app:ClusterDefinition}
To explore the influence of cluster definitions, we constructed clusters between four strongly coupled pigments in photosystem I (PSI). We use an iterative, three-step algorithm designed to ensure that strongly interacting pigments are preferentially included inside the same cluster. 
\begin{enumerate}
    \item \textbf{Step 1:} Locate the largest coupling element ($V_{i,j}$) among the pigments not yet assigned to a cluster. Pigments $(i,j)$ for the nucleus of a new cluster. 
    \item \textbf{Step 2:} Locate the largest coupling element to either pigment in the new cluster to another pigment not yet assigned to a cluster ($V_{m,j}$ or $V_{i,m}$). Add this pigment to the new cluster: $(i,j,m)$. 
    \item \textbf{Step 3:} Locate the largest coupling element involving a pigment in the new cluster with another pigment ($n$) not yet assigned to a cluster. Add this pigment to the new cluster: $(i,j,m,n)$.
    \item \textbf{End Condition:} If any pigments remain unassigned then return to step 1 and define a new cluster.
\end{enumerate}
This algorithm does not guarantee that all strong couplings are contained within a cluster, but it does ensure that clusters nucleate around strong coupling elements.  

\section*{References}
	\bibliography{Adaptive-Abs-JCP}
	\bibliographystyle{apsrev}
\end{document}